\pdfoutput = 1
\documentclass[journal=nalefd, manuscript=letter]{achemso}
\usepackage{graphicx}
\usepackage{graphics}
\usepackage{amsmath}
\usepackage{amssymb}
\usepackage{amsfonts}
\usepackage{dcolumn}
\usepackage{dsfont}
\usepackage{latexsym}
\usepackage{rotating}
\usepackage{array}
\usepackage{makecell}
\usepackage{color}
\usepackage{latexsym}
\usepackage{bbm}
\usepackage{physics}
\usepackage{subfigure}
\usepackage{float}
\usepackage{epsfig}
\usepackage{epsf}
\usepackage{psfrag}
\usepackage{bm}
\usepackage{amsthm}
\usepackage{eucal}
\usepackage{mathrsfs}
\usepackage{tabularx}
\usepackage{url}
\usepackage{braket}
\usepackage{epstopdf}
\usepackage{color} % use colored text in latex

\usepackage{hyperref}
\hypersetup{
colorlinks=true,
        final=true,
        linkcolor=black,
        citecolor=black,
        filecolor=black,
        urlcolor=black,
}
\title{Charge Density Waves and In-Gap States in Coupled Chains and Moir\'e Superlattices}

\author{Muhammad Akram}
\affiliation{Department of Physics, Arizona State University, Tempe, AZ 85287, USA}
\affiliation{Department of Physics, Balochistan University of Information Technology, Engineering and Management Sciences (BUITEMS), Quetta 87300, Pakistan}
\email{makram3@asu.edu}

\author{Onur Erten}
\affiliation{Department of Physics, Arizona State University, Tempe, AZ 85287, USA}
\email{onur.erten@asu.edu}

\begin{document}


\begin{abstract}
Charge density waves (CDWs) arise from the interplay between electrons and lattice distortions and are a hallmark of low-dimensional systems. Here we study coupled Su-Schrieffer-Heeger chains, both lattice-matched ladders and lattice-mismatched moir\'e superlattices, and find a rich set of CDW phases together with symmetry-protected bound states. In lattice-matched chains, the competition between electron-phonon coupling and interchain tunneling drives a sequence of CDW phases, including staggered dimers, trimers, and incommensurate states. These phases are generally gapped and host bound states protected by crystalline symmetries rather than chiral symmetry, and are therefore not pinned to zero energy. Lattice mismatch further reconstructs these phases into long-period supermoir\'e CDWs, with an ordered array of defects, pinned to the moir\'e length scale, that host localized electronic states. Interchain tunneling and lattice mismatch thus emerge as natural control parameters for CDW order and bound states in one-dimensional chains.
\end{abstract}

%%%%%%%%%%%%%%%% INTRODUCTION 
Charge density waves (CDWs), which arise from the interplay between electronic and lattice degrees of freedom, are among the most extensively studied collective instabilities in low-dimensional quantum materials~\cite{GrunerReview,GorkovGruner1989,Gruner1994DensityWaves, Monceau2012,Pouget2016}. In one dimension, CDW formation is captured by the Peierls instability, in which electron-phonon coupling drives a spontaneous lattice distortion that enlarges the unit cell and opens a gap at the Fermi level~\cite{Peierls}. The Su-Schrieffer-Heeger (SSH) model provides the minimal description of this physics and famously hosts topological bound states in its dimerized phase~\cite{SSH1,SSH2}.

The discovery of correlated phases in twisted bilayer graphene established moir\'e superlattices as a versatile way to create and control quantum phases~\cite{Cao2018, Bistritzer_PNAS2011}. This approach has since produced unconventional superconductivity, correlated insulators, and Wigner crystals across graphene and semiconducting transition metal dichalcogenide bilayers~\cite{Andrei_NatMat2020, Mak_NatNano2022}. Most of these systems are weakly correlated on their own, and the moir\'e potential drives them into ordered phases. A less explored question is what happens when the constituent layers already host their own broken-symmetry order. Twisted van der Waals magnets are one active example, where the moir\'e periodicity gives rise to a variety of rich magnetic phase diagrams~\cite{Hejazi_PNAS2020, Hejazi_PRB2021, Akram_PRB2021, Tong_PRR2021, Akram_NanoLett2021, Akram_NanoLett2024, Akram_NanoLett2025, Xu2022, Wong2026, Song2021, Yang2024}.

Charge density waves are a natural second example, and their interplay with moir\'e modulation has only recently been examined. Experiments on twisted metallic dichalcogenides reveal CDW domains pinned by the moir\'e pattern~\cite{Zhao_NatMat2022}, and theory has described how a monolayer CDW of fixed period is accommodated within a moir\'e domain structure~\cite{Goodwin_JPCM2022}. In these studies the CDW is a pre-formed electronic order of fixed period, and the moir\'e enters only through the strain landscape of the reconstructed lattice, which traps or interrupts it.

More broadly, the accommodation of a density wave to a competing periodicity is a classic problem: a CDW incommensurate with its host relieves the mismatch through a periodic array of discommensurations, domain walls that separate locally commensurate regions~\cite{McMillan1976, PokrovskyTalapov1979, Bak1982}. Two features set the present problem apart. First, the CDW is of Peierls origin, so the distortion that opens the gap is the same lattice field that must accommodate the mismatch. Its periodicity, therefore, cannot be imposed independently and relaxed afterward; rather, it emerges self-consistently from the competition between electronic and elastic energies. Second, the mismatch is a geometric moir\'e between two coupled chains rather than between a single CDW and its substrate, so the discommensurations are not phase slips of a scalar order parameter but are tied to the local interchain stacking as it evolves through AB, AA, and BA registry. Solving the displacement field self-consistently in this setting produces supermoir\'e CDWs whose periods are multiples of the moir\'e period. These states host a tunable array of stacking-locked defects, and electronic bound states are attached to these defects with a spacing set by the moir\'e scale. A recent study of one-dimensional moir\'e ladders treated incommensurate CDWs of excitonic origin and their phason dynamics.~\cite{Mellado_APSOpSci2026, Mellado_PRB2026} However, the self-consistent electron–phonon problem in coupled chains remains unexplored.

Motivated by this, we study a two-leg SSH model at half filling. We treat the phonons as classical and determine the ground state distortion and charge density by Monte Carlo simulations. Our main results are as follows. (i) For lattice-matched chains, the competition between interchain tunneling and electron-phonon coupling produces a sequence of CDW phases, including staggered dimers, trimers, and incommensurate states with enlarged unit cells. (ii) These phases are gapped and can host boundary modes protected by crystalline symmetries; although long-range interchain tunneling breaks chiral symmetry, these modes are not pinned to zero energy. (iii) A lattice mismatch reorganizes these phases into supermoir\'e CDWs whose period is a multiple of the moir\'e period, with an ordered, tunable array of stacking-locked defects. These defects bind localized electronic states whose spacing is set by the moir\'e length scale.

\begin{figure}
\center
\includegraphics[width=1\textwidth]{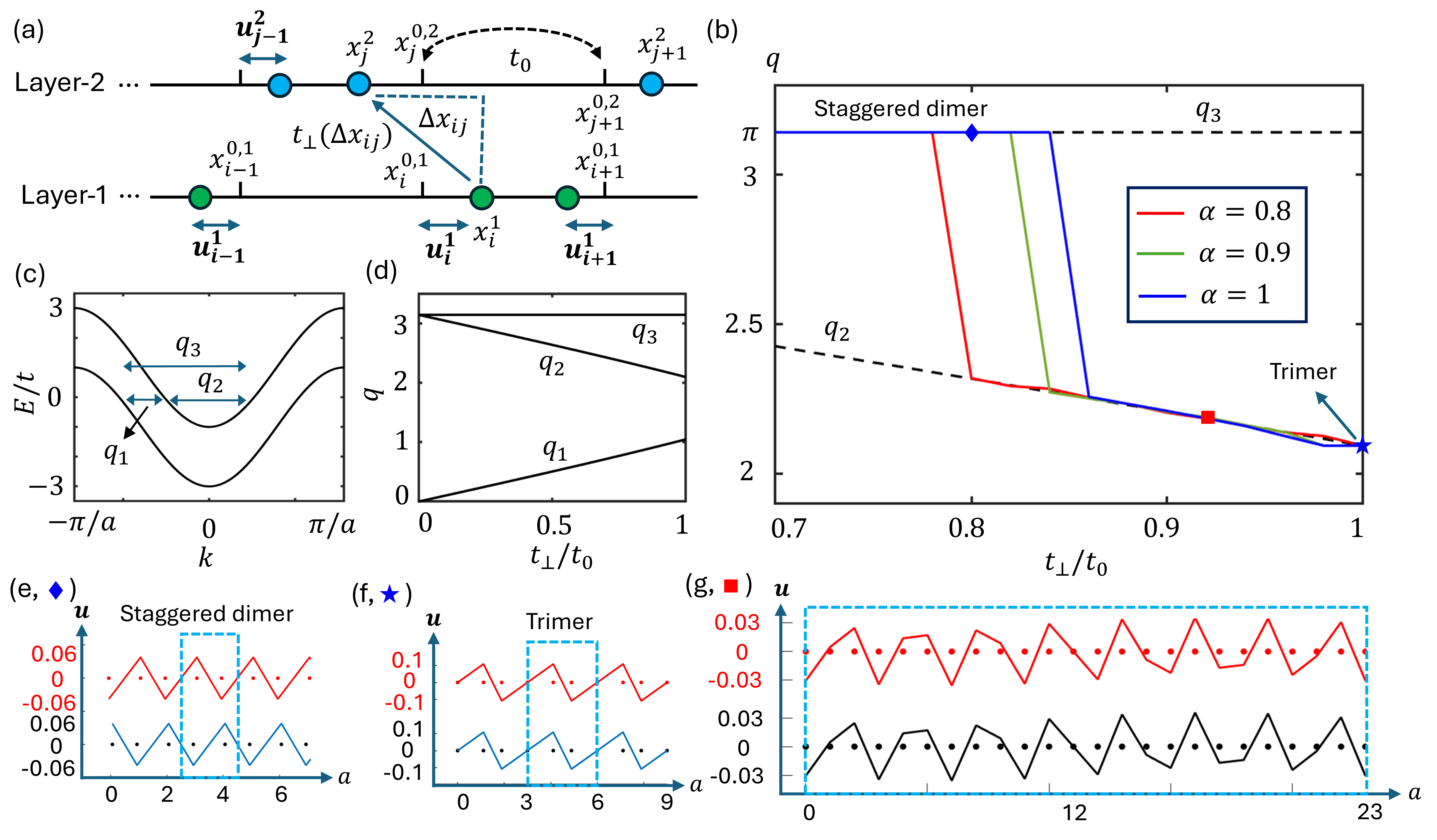} 
\caption{Emergent charge density wave (CDW) phases in lattice-matched chains. (a) Illustration of the model highlighting intrachain tunneling $t_0$, interchain tunneling $t_\perp$, lattice distortions $u_i$, and in-plane displacement $\Delta x_{ij}=x_i^1-x_j^2$ between the atoms $i$ and $j$ in the two chains. (b) Ground-state phase diagram in the $(t_\perp/t_0,\alpha/t_0)$ plane, revealing a wide range of CDW phases, including the staggered-dimer and trimer phases. (c) Energy spectrum $E(k)$ of coupled chains with nearest-neighbor interchain tunneling, in the absence of dimerization. The characteristic wave vectors $q_{1}$, $q_{2}$, and $q_{3}=q_{1}+q_{2}$ are defined from the corresponding Fermi points, as indicated. (d) Evolution of the characteristic wave vectors $q_{1}$, $q_{2}$, and $q_{3}$ as a function of the interchain tunneling $t_{\perp}$. (e) Staggered-dimer phase with a two-site unit cell, where the dimers of one chain are centered between adjacent dimers of the other chain, at $t_{\perp}/t_{0}=0.8$ and $\alpha/t_{0}=1$.  (f) Trimer ($\lambda_3$) phase of unit cell size of three sites with one dimer and one site unshifted at $t_\perp / t_0=1$ and $\alpha/t_0=1$. The light blue dashed rectangular box represents the unit cell. (g) $L=23a$ CDW phase with $t_\perp / t_0=0.92$ and $\alpha/t_0=0.8$.}
\label{Fig:1}
\end{figure}

%%%%%%%%%%%%%%%% HAMILTONIAN %%%%%%%%%%%%%%%%%%%%%%%%%
We consider
$\mathcal{H} = \mathcal{H}_{\mathrm{intra}}+ \mathcal{H}_{\mathrm{inter}}$,
where $\mathcal{H}_{\mathrm{intra\,(inter)}}$ describes the intra (inter) chain Hamiltonian,
\begin{eqnarray}
\label{Eq:Gnrl_Hmlt_1}
\mathcal{H}_{\mathrm{intra}}&=&-\sum_{i,\beta}t(u_{(i+1)\beta}-u_{i\beta})(c_{(i+1)\beta}^\dagger c_{i\beta}+ {\rm H.c.})+\frac{K}{2}\sum_{i,\beta} (u_{(i+1)\beta}-u_{i\beta})^2,\\
\mathcal{H}_{\mathrm{inter}}&=& -\sum_{ij} t_\perp(x_{i}^1-x_{j}^2)\,(c_{i1}^{\dagger} c_{j2}+{\rm H.c.}).
\end{eqnarray}
Here, $\beta = 1, 2$ is the chain index, $u_i^\beta$ is the displacement of the $i^{\rm th}$ ion in chain $\beta$ from its equilibrium position $x_i^{0,\beta}$, so that its in-plane position is $x_i^\beta = x_i^{0,\beta} + u_i^\beta$, and $K$ is the spring constant of the phonon mode (Fig.~\ref{Fig:1}(a)). The intrachain hopping depends on the change in bond length $u_{(i+1)\beta}-u_{i\beta}$, whereas the interchain hopping depends on the in-plane separation $x_i^1 - x_j^2 = (x_i^{0,1}-x_j^{0,2}) + (u_i^1 - u_j^2)$ between sites on the two chains. This separation carries the lattice mismatch through the equilibrium offset $x_i^{0,1}-x_j^{0,2}$ and the distortion through the displacements. For lattice-matched chains the paired sites are aligned, $x_i^{0,1}=x_j^{0,2}$, so that $x_i^1 - x_j^2 = u_i^1 - u_j^2$. The two hopping amplitudes are given by
\begin{eqnarray}
t(u_{(i+1)\beta}-u_{i\beta}) &=& t_0 - \alpha\,(u_{(i+1)\beta}-u_{i\beta}), \\
t_\perp(x_{i}^1-x_{j}^2) &=& t_\perp\exp\!\left[-\gamma\left(\sqrt{(x_{i}^1-x_{j}^2)^2+1}-1\right)\right],
\end{eqnarray}
where $\alpha$ sets the electron-phonon coupling strength: as neighboring sites move closer or farther apart, the hopping amplitude increases or decreases, respectively. The interchain tunneling depends on the in-plane separation, with the interchain spacing fixed to one lattice constant (the $+1$ under the square root); it is maximal, $t_\perp$, when the two sites are aligned across the chains ($x_i^1=x_j^2$) and decays exponentially with separation, following previous studies~\cite{Greiner2002,Schneider2012,Jo2012}. Throughout, we set $\gamma=5$, so that the interchain tunneling is short-ranged, becoming negligible beyond one lattice constant and effectively vanishing by a separation of $\sim 2a$. Our Hamiltonian is readily generalized to moir\'e superlattices by giving the two chains different lattice constants.

We explore the phase diagram by treating the displacements $u_i^\beta$ as classical variables and performing classical Monte Carlo (MC) simulations with simulated annealing. For each displacement configuration we diagonalize the electronic Hamiltonian and evaluate the total energy as the sum of the occupied electronic levels and the elastic energy $\tfrac{K}{2}\sum_{i,\beta}(u_{(i+1)\beta}-u_{i\beta})^2$. The displacements are then updated by the Metropolis algorithm and relaxed toward the minimum-energy configuration. The system is annealed in three stages of ten steps each, $T/t_0 = 10^{-4}\!\to\!10^{-6}$, $10^{-6}\!\to\!10^{-7}$, and $10^{-7}\!\to\!10^{-8}$, with $2000\times L$ MC steps at each temperature. The procedure is repeated for $48$ independent random initial configurations, and the lowest-energy result is taken as the ground state. We use periodic boundary conditions, with system sizes $L=20a$ to $66a$ for the lattice-matched chains and $L>60a$ for the moir\'e superlattices to reduce finite-size effects.

We begin our analysis with the ground-state phase diagram of the coupled chains as a function of interchain tunneling and electron-phonon coupling. Fig.~\ref{Fig:1}(b) presents this phase diagram in the $(t_{\perp},\alpha)$ plane. Over a broad range of interchain tunneling, the system stabilizes a Peierls-type staggered-dimer phase, in which the dimers of one chain are centered between adjacent dimers of the other, giving a two-site unit cell of length $2a$, as illustrated in Fig.~\ref{Fig:1}(e). A natural competitor is the AA-stacked dimer, in which the dimers of the two chains sit directly on top of one another. Because this alignment minimizes the interchain distance and therefore maximizes the tunneling, the dimerization gap is strongly suppressed and closes near $t_{\perp}/t_0 = 0.25$, making the AA-stacked dimer energetically unfavorable. In the staggered configuration, the interchain tunneling is weaker, so the gap is reduced gradually and remains open across the full range of $t_\perp$. Consequently, the staggered-dimer phase is more stable than the AA-stacked configuration. The evolution of the gap for both configurations is shown in Fig.~1 of the Supporting Information.

With increasing interchain tunneling, the staggered-dimer phase transitions into a sequence of long-period, incommensurate CDW phases with enlarged unit cells, setting in beyond $t_\perp/t_0 \approx 0.80$, $0.84$, and $0.86$ for $\alpha/t_0 = 0.8$, $0.9$, and $1.0$, respectively. A representative example, an $L=23a$ phase at $t_\perp/t_0=0.92$ and $\alpha/t_0=0.8$, is shown in Fig.~\ref{Fig:1}(g). Upon further increasing $t_\perp$, the system enters the trimer phase, realized at $t_\perp/t_0=1$ for all three values of $\alpha$. This phase is characterized by a three-site unit cell consisting of one dimer and one undistorted site, as illustrated in Fig.~\ref{Fig:1}(f).

To understand the origin of these phases, we track the dominant ordering wave vector as a function of $t_\perp$ [Fig.~\ref{Fig:1}(d)]. For nearest-neighbour AA-stacked coupled chains, the bonding and antibonding bands produce four Fermi points, and a CDW can open a gap at any of the wave vectors connecting them, $q_1$, $q_2$, and $q_3 = q_1 + q_2$ [Fig.~\ref{Fig:1}(c)]. At half filling the Fermi momenta shift with $t_\perp$ as the two bands split, so $q_1$ and $q_2$ evolve, but their sum is fixed and $q_3 = \pi$ is independent of $t_\perp$. A Peierls distortion lowers the electronic energy most when its wave vector connects states at the Fermi level. At $t_\perp = 0$ the two bands are degenerate and this condition is met at $q_3 = \pi$, stabilizing the two-site staggered dimer. As $t_\perp$ grows and the bands split, the dominant instability moves to $q_2$. Because $q_2$ then decreases continuously away from $\pi$, it generates the sequence of long-period, incommensurate CDWs, locking into the three-site trimer when $q_2$ reaches the commensurate value $2\pi/3$ near $t_\perp/t_0 = 1$. Although the full model includes longer-range interchain tunneling, the extracted ordering vectors closely follow this nearest-neighbour reference, confirming that it captures the essential CDW instability. For $\alpha/t_0 = 0.8$ and $0.9$ the ordering vector stays locked to $q_2$ up to $t_\perp/t_0 = 1$, whereas for $\alpha/t_0 = 1.0$ the trimer sets in slightly below $t_\perp/t_0 = 1$. We restrict our analysis to $t_\perp \le t_0$, the regime relevant to realistic coupled chains.

\begin{figure}
\center
\includegraphics[width=0.5\textwidth]{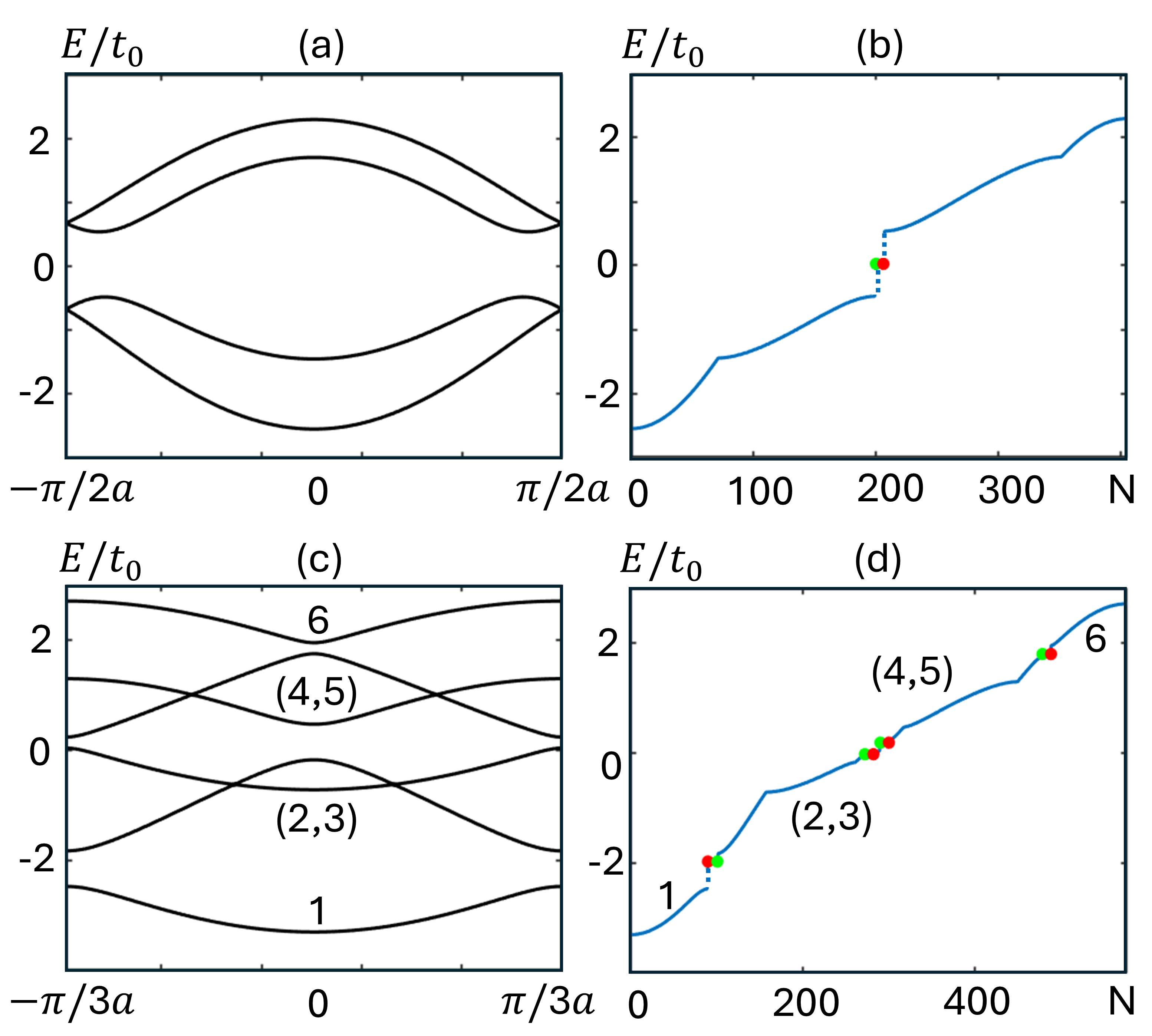} 
\caption{Band structure and real-space (open-boundary) energy spectrum for staggered-dimer and trimer phases. 
(a,b) Staggered-dimer at $t_\perp/t_0 = 0.5$, $\alpha = 1$, showing bulk bands and localized edge in-gap states at the opposite boundaries (red and green). 
(c,d) Trimer $(\lambda_3)$ at $t_\perp/t_0 = 1$, $\alpha = 1$, with edge states localized at opposite boundaries (green and red circles).}
\label{Fig:2}
\end{figure}

To further characterize the CDW phases, we focus on the staggered-dimer and trimer phases, analyzing their bulk electronic structure and boundary spectra. Figure~\ref{Fig:2} presents the band structures together with the open-boundary energy spectra; both phases are fully gapped and host localized in-gap states at the edges (red and green circles). In the standard SSH model such edge modes are protected by chiral symmetry (BDI class)~\cite{su1979solitons, ryu2010topological}. Here, however, the long-range interchain tunneling introduces hopping within the same sublattice, breaking the bipartite structure and hence chiral symmetry, so the model cannot be classified within the BDI framework. We therefore turn to its crystalline symmetries. All phases in Fig.~\ref{Fig:1} preserve the mirror symmetries about the $x$- and $y$-axes and inversion symmetry, with one exception: the staggered-dimer phase is not invariant under a pure $M_y$ mirror but instead preserves the nonsymmorphic symmetry $T_aM_y$, where $T_a$ is a translation by one lattice site along the chain. These crystalline symmetries protect a nontrivial bulk topology whose boundary states, unlike the chiral-symmetry-protected zero modes of the SSH model, are not pinned to zero energy~\cite{Fu2011,Bradlyn2017,Chiu2016,Alexandradinata2014,Lin2025}.

We begin the topological characterization with the inversion invariant. For a one-dimensional inversion-symmetric insulator it is given by the product of the inversion eigenvalues of the occupied bands at the inversion-invariant momenta $k=0$ and $k=\pi$,
\begin{equation}
\nu_{\mathrm{inv}}=\prod_{n\in\mathrm{occ}}
\xi_n(0)\,\xi_n(\pi),
\end{equation}
where $\xi_n(k)=\pm1$ is the inversion eigenvalue of the $n^{\text{th}}$ occupied band~\cite{Hughes2011}. For the staggered-dimer phase we obtain $\nu_{\mathrm{inv}}=-1$, identifying it as a nontrivial inversion-protected topological phase. This is independently confirmed by a Wilson-loop analysis~\cite{Alexandradinata2014}, with details in the Supporting Information.

For the trimer phase, both the inversion invariant ($\nu_{\mathrm{inv}}=1$) and the Wilson-loop phase ($\theta=0$) identify the occupied subspace as topologically trivial. Nevertheless, the open-boundary spectrum in Fig.~\ref{Fig:2}(d) shows several localized edge states within the bulk gaps. Their origin lies in the combined time-reversal and inversion symmetry ($\mathcal{PT}$), under which each isolated band, or isolated group of bands, carries an integer-valued inversion winding number~\cite{Lin2025},
\begin{equation}
\nu=\frac{1}{\pi i}\int_{0}^{2\pi}dk\,
\frac{\langle u_k|\partial_k(\hat{P}_3-\hat{P}_1)|u_k\rangle}
{\langle u_k|(\hat{P}_1+\hat{P}_3)|u_k\rangle},
\end{equation}
where $\hat{P}_{\sigma}=\sum_k |k,\sigma\rangle\langle k,\sigma|$ projects onto sublattice $\sigma=1,2,3$. This gauge-invariant $\mathbb{Z}$ invariant obeys a bulk-boundary correspondence, with $2\nu$ edge states associated with a band (or band subspace) of winding number $\nu$. The lowest and highest bands (1 and 6) are spectrally isolated, each with $\nu=1$, predicting two edge states, in agreement with the spectrum. The middle pairs $(2,3)$ and $(4,5)$ touch, so the winding is ill-defined for the individual bands, but each two-band subspace carries a quantized winding $\nu=3$ and hence six edge states, again consistent with Fig.~\ref{Fig:2}(d). Thus, although the occupied subspace is topologically trivial, the isolated bands and composite subspaces carry a nontrivial $\mathbb{Z}$ topology protected by $\mathcal{PT}$ symmetry, accounting for the observed edge states.

\begin{figure}
\center
\includegraphics[width=\textwidth]{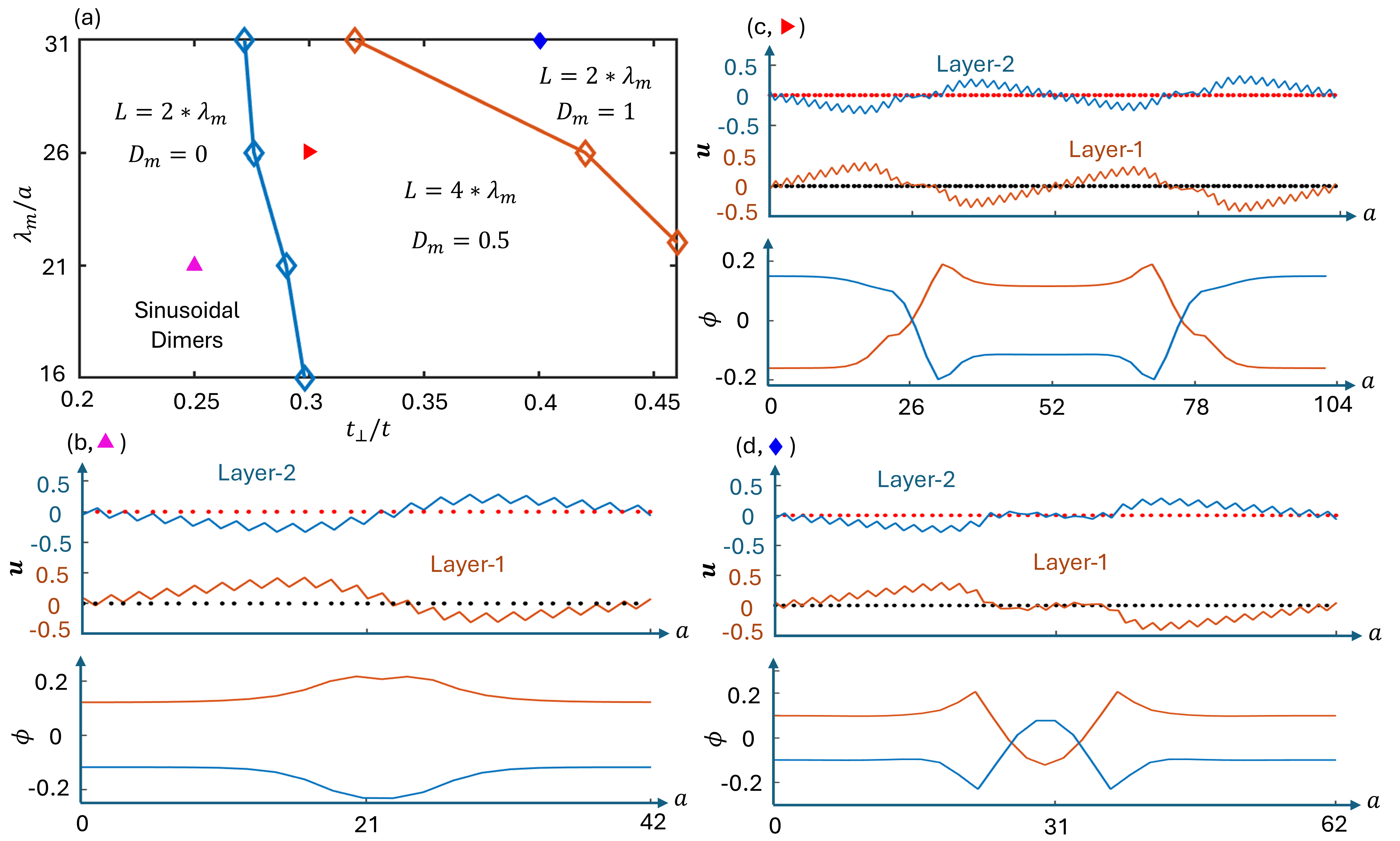} 
\caption{Emergent charge density wave (CDW) phases in mismatched coupled chains. (a) Phase diagram of the CDW phases as a function of the interchain tunneling
$t_\perp/t_0$ and moir\'e period $\lambda_m/a$, showing supermoir\'e phases with emergent periodicities $L = 2\lambda_m$, $4\lambda_m$, and $2\lambda_m$. Here, $D_m$ denotes the defect density within a moir\'e unit of length
$\lambda_m$. (b) Supermoir\'e phase with $L = 2\lambda_m$ and $D_m = 0$ at $t_\perp/t_0 = 0.25$ and $\lambda_m = 21a$. (c) Supermoir\'e phase with $L = 4\lambda_m$ and $D_m = 0.5$ at $t_\perp/t_0 = 0.3$ and $\lambda_m = 26a$. (d) Supermoir\'e phase with $L = 2\lambda_m$ and $D_m = 1$ at $t_\perp/t_0 = 0.4$ and $\lambda_m = 31a$. In (b-d), the top panels show the site displacements relative to the mean lattice position, while the bottom panels show the corresponding CDW order parameter, $\phi = u_A - u_B$.}
\label{Fig:3}
\end{figure}

We now turn to the effect of a lattice mismatch, which introduces a moir\'e modulation of the interchain stacking. Figure~\ref{Fig:3}(a) shows the phase diagram as a function of the interchain tunneling $t_\perp$ and the moir\'e period $\lambda_m$ at $\alpha/t_0=0.8$. We focus on the weak-tunneling regime $t_\perp/t_0\lesssim0.45$, where the matched chains form the staggered-dimer phase. The mismatch makes the interchain stacking vary along the chains, driving the system through a sequence of supermoir\'e phases of increasing period. At small interchain tunneling, a supermoir\'e phase of period $L=2\lambda_m$ emerges, with a long-range sinusoidal modulation of the displacements [Fig.~\ref{Fig:3}(b)]. The staggered-dimer order survives only over part of each moir\'e supercell, and it grows with $t_\perp$. Near $t_\perp/t_0=0.1$, it fills an entire moir\'e cell. Elsewhere, the stacking evolves continuously from AB to AA to BA, and an extra dimer is inserted into the chain with the smaller lattice constant. Within each moir\'e cell the dimers of the two chains shift in opposite directions, and this pattern reverses from one cell to the next, causing the displacement field $u$ to change sign between adjacent cells [Fig.~\ref{Fig:3}(b)]. The order parameter $\phi_i = u_{iA}-u_{iB}$, by contrast, keeps a fixed sign along each chain and opposite signs in the two chains. This out-of-phase dimerization is the defining feature of the $L=2\lambda_m$ phase, and the amplitude of the modulation grows with $t_\perp$. The corresponding energetics are shown in Fig.~2(a,b) of the Supporting Information, which plots the energy per unit cell as a function of interchain displacement $\Delta x$. At $t_\perp/t_0=0.1$ and $0.2$, the energy is minimized for staggered-dimer stacking ($\Delta x=0$) and maximized for AA stacking ($\Delta x=1$). Thus, in the presence of lattice mismatch, the system favors staggered-dimer stacking while avoiding AA stacking.

Increasing $t_\perp$ drives the system into an $L=4\lambda_m$ supermoir\'e phase with a periodic array of defects, at a defect density $D_m=0.5$ (one defect per two moir\'e cells) [Fig.~\ref{Fig:3}(c)]. These defects are discommensurations of the staggered-dimer order: unshifted intradimer sites where $\phi_i$ crosses zero and changes sign. They align vertically across the two chains and sit adjacent to the AA-stacked regions, which they enlarge. Consequently, the phase consists of two staggered-dimer domains separated by variable-stacking regions containing the defects. The energetics are shown in Figs.~2(c,d) of the Supporting Information. At $t_\perp/t_0=0.3$ and $0.4$, staggered-dimer stacking ($\Delta x=0$) remains the lowest-energy configuration. However, unlike in the weak-tunneling regime ($t_\perp/t_0=0.1$ and $0.2$), AA stacking is no longer the highest-energy configuration. The phase is stabilized by the competition between the interchain energy gained by enlarging the staggered-dimer and AA-stacked regions and the elastic cost of forming defects. It persists over a broad range of $t_\perp$, particularly at shorter moir\'e periods. With further increase of $t_\perp$, and at large moir\'e periods, the system transitions to an $L=2\lambda_m$ phase with defect density $D_m=1$ (two defects per two moir\'e cells). The defects are separated by expanded AA-stacked regions that lower the interchain energy. Although the second defect carries an energy cost, the strong interchain tunneling stabilizes this configuration.

\begin{figure}
\center
\includegraphics[width=1\textwidth]{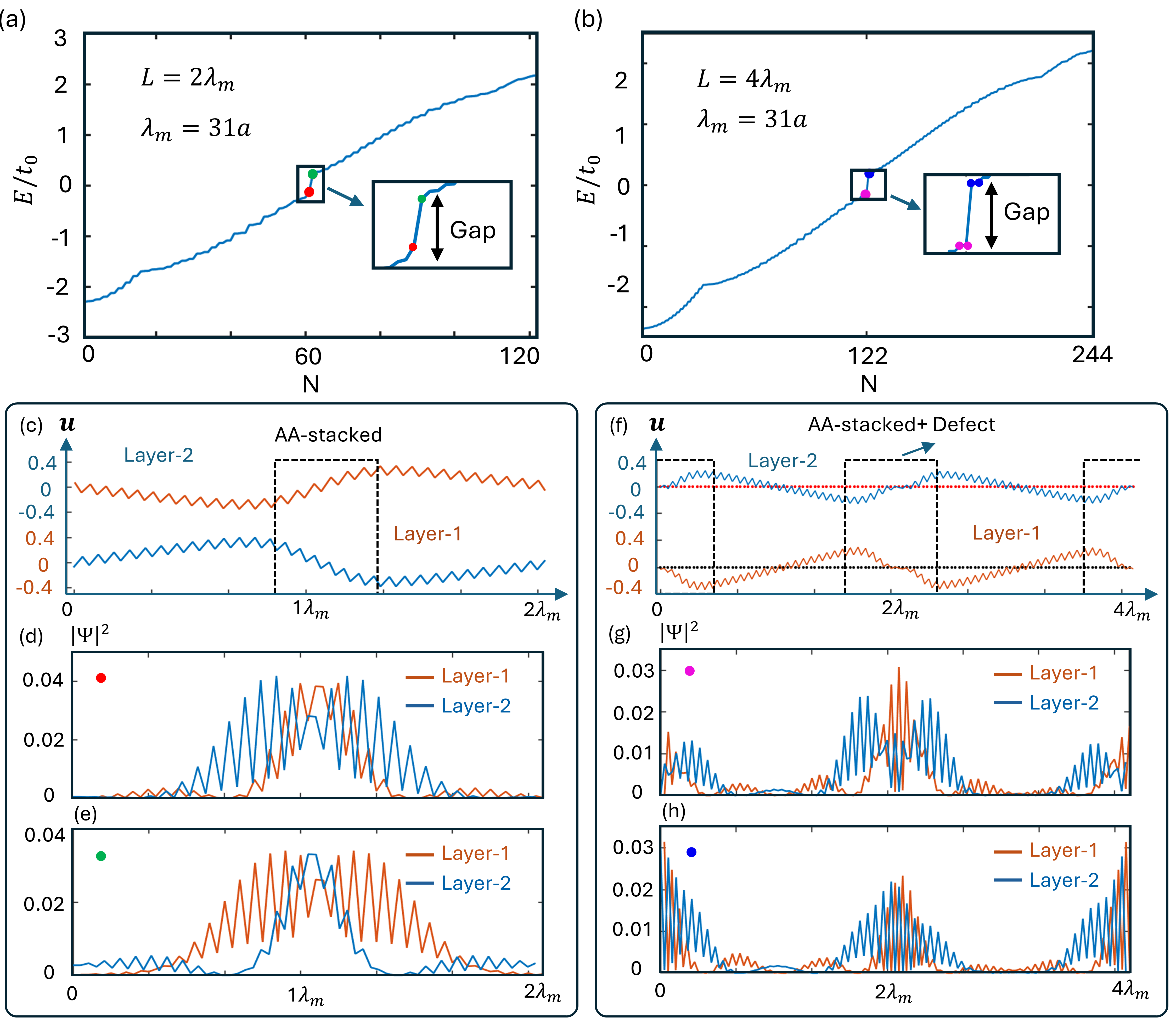} 
\caption{Electronic band structure and real-space probability density profiles of super-moir\'e CDW phases in mismatched coupled chains. (a) Band structure of the $L = 2\lambda_m$ super-moir\'e phase at $t_\perp/t_0 = 0.25$ and $\lambda_m = 31a$. The spectrum is gapped, with red markers indicating states near the band edges corresponding to modes localized at defect sites. (b) Band structure of the $L = 4\lambda_m$ super-moir\'e phase at $t_\perp/t_0 = 0.3$ and $\lambda_m = 31a$, also showing a gapped spectrum with states near the band edges exhibiting similar spatial localization. (c) Spatial profile of the $L = 2\lambda_m$ super-moir\'e phase. (d, e) Probability density of representative states (62 and 63), showing localization at the defect position $x = 1\lambda_m$. (f) Spatial profile of the $L = 4\lambda_m$ super-moir\'e phase. (g, h) Probability density of states (122 and 123), showing localization at the defect positions $x =0, 2\lambda_m$.}
\label{Fig:4}
\end{figure}

To examine the electronic structure of these phases, we compute their spectra [Fig.~\ref{Fig:4}(a,b)]. Both the $L=2\lambda_m$ and $L=4\lambda_m$ supermoir\'e phases are fully gapped and host localized in-gap states (colored dots). In the $L=2\lambda_m$ phase ($t_\perp/t_0=0.25$) these states live in the variable-stacking region separating neighboring staggered-dimer domains, where the local stacking changes continuously from AB to AA to BA [Fig.~\ref{Fig:4}(d,e)]. To resolve their origin and extent, we enlarge the system from $L=62a$ to $L=602a$ (Fig.~3 of the Supporting Information). The resulting dense set of in-gap states is tightly localized around the AA-stacked region near the gap center, where the AA-stacked bilayer is nearly gapless at $t_\perp/t_0\approx0.25$. At energies away from the gap center, the states spread progressively into the neighboring stacking regions and eventually merge with the delocalized bulk states at the gap edges. The spatial extent of the in-gap states thus follows the stacking-dependent local gap across the moir\'e cell.

The same picture applies to the $L=4\lambda_m$ phase, where the staggered-dimer domains are separated by variable-stacking regions containing the intradimer defects near the AA-stacked regions [Fig.~\ref{Fig:4}(f)]. In an isolated SSH chain, such defects would bind chiral-symmetry-protected states. Here, long-range interchain tunneling breaks chiral symmetry. Nevertheless, because the AA-stacked chains are locally gapless at this tunneling strength, the states remain localized near the AA regions [Fig.~\ref{Fig:4}(g,h)]. As their energies move away from the gap center, they extend further into the neighboring stacking regions. The phase $L=2\lambda_m$, $D_m=1$ exhibits a similar behavior. Its staggered-dimer domains are separated by an enlarged AA-stacked region bounded by two defects. This locally gapless region also binds in-gap states, whose spatial extent depends on the local stacking,  with the moir\'e period acting as a tunable handle on their location.

A direct solid-state realization of the coupled-chain geometry studied here is
provided by indium atomic wires on Si(111), where a metallic array of quantum
chains undergoes a spontaneous Peierls transition into a charge density wave
ground state built from two mutually coupled, dimerized chains that host 
bound states~\cite{Yeom1999, Cheon2015}. This system realizes the same class of
electron-phonon physics as our model, with the lattice distortion determined
self-consistently rather than imposed externally. The resulting spectra,
topological invariants, and edge modes can be independently probed in
programmable platforms that emulate the coupled-chain Hamiltonian, including
ultracold-atom SSH lattices in which the Zak phase is measured
directly~\cite{Atala2013}, photonic lattices that host multiple topological edge
states beyond the single-mode SSH limit~\cite{Zhang2019}, acoustic metamaterials
built from coupled SSH chains~\cite{Guo2024}, and topoelectric-circuit
implementations of the two-leg SSH ladder~\cite{Kuanr2026}. In these synthetic
systems the intrachain dimerization and interchain tunneling are set by design,
so the staggered-dimer, trimer, and longer-period spectra identified here can be
mapped out directly, while the self-consistent selection of the ordering vector
remains specific to the electron-phonon problem. On the moir\'e side,
one-dimensional moir\'e chains with strongly localized electronic states have
been realized in twisted bilayer WSe$_2$~\cite{Ren2024}, and twisted bilayers of
anisotropic materials such as GeSe, together with twisted rectangular lattices,
provide a route to long-wavelength one-dimensional moir\'e
structures~\cite{Kennes2020, An2025}. A particularly direct realization of
lattice-mismatched coupled one-dimensional chains is provided by incommensurate
double-walled carbon nanotubes, whose mismatched shells form a one-dimensional
moir\'e crystal~\cite{Koshino2015}, and by natural misfit compounds such as
(BiSe)$_{1+\delta}$NbSe$_2$, in which incommensurate stacking produces
quasi one-dimensional electronic states in a charge density wave
host~\cite{Chikina2022}. Although the localization in these moir\'e and misfit
systems originates from the interlayer potential and registry rather than a
self-consistent Peierls distortion, they point toward the mismatched coupled
chains studied here, in which the displacement field itself accommodates the
lattice mismatch.

In summary, we have shown that in coupled Peierls chains the competition between electron-phonon coupling and interchain tunneling stabilizes a sequence of CDW phases, from staggered dimers through incommensurate states to trimers. Although the long-range interchain tunneling breaks chiral symmetry, these phases host boundary states protected by crystalline symmetries and characterized by inversion and $\mathcal{PT}$ invariants, with energies that need not lie at zero. A lattice mismatch reorganizes the CDW into supermoir\'e phases in which the dimer order accommodates the mismatch through an ordered array of discommensurations, and the resulting defects bind localized in-gap states whose spacing is set by the moir\'e period. Interchain tunneling and lattice mismatch thus act as control parameters for CDW order and bound states in one-dimensional chains.

\section{Associated Content}

\noindent {\bf Supporting Information}

\noindent {Details of the Wilson Loop Analysis, evolution of the energy gap in dimer phases and evolution of the energy with interchain displacement (pdf)}.

\section{Author Information}
\noindent {\bf Corresponding Author:}
Muhammad Akram -- Email: makram3@asu.edu \\ Onur Erten -- Email: onur.erten@asu.edu

\noindent {\bf Notes:} The authors declare no competing financial interest.

\section{Acknowledgements}
We thank M.~\"O.~Oktel and Pedro M.~C\^onsoli for fruitful discussions. Part of this work, the topological characterization of the charge density wave and moir\'e phases, analysis of the results, was supported by the U.S. Department of Energy, Office of Science, Office of Basic Energy Sciences, Materials Sciences and Engineering Division under Award Number DE-SC0025247 (OE).

%\begin{center}
%    \includegraphics[width=3.25in]{TOC_fig.png}
%    %Table of content graphic
%\end{center}

%\bibliography{references}

\providecommand{\latin}[1]{#1}
\makeatletter
\providecommand{\doi}
  {\begingroup\let\do\@makeother\dospecials
  \catcode`\{=1 \catcode`\}=2 \doi@aux}
\providecommand{\doi@aux}[1]{\endgroup\texttt{#1}}
\makeatother
\providecommand*\mcitethebibliography{\thebibliography}
\csname @ifundefined\endcsname{endmcitethebibliography}  {\let\endmcitethebibliography\endthebibliography}{}

\end{document}

\maketitle

\section{Details of the Wilson Loop Analysis}\label{sec:wilson}
The Wilson loop over the occupied bands is defined as
\begin{equation}
\mathcal{W}
=
\mathcal{P}
\exp\!\left[
i\oint_{\mathrm{BZ}}
\mathbf{A}(k)\,dk
\right],
\end{equation}
where $\mathcal{P}$ denotes path ordering along the Brillouin zone, and $\mathbf{A}_{mn}(k)=i\langle u_m(k)|\partial_k u_n(k)\rangle$ is the non-Abelian Berry connection. The eigenvalues of $\mathcal{W}$ are of the form $e^{i\theta_j}$, where $\theta_j$ are the Wilson-loop phases~\cite{Alexandradinata2014}. For the staggered-dimer phase, the Wilson-loop spectrum contains an eigenvalue of $-1$, corresponding to a Wilson-loop phase $\theta=\pi$ (mod $2\pi$), fully consistent with the inversion topological invariant and confirming the nontrivial topology of the staggered-dimer phase. For the trimer phase, by contrast, the Wilson-loop phase is $\theta=0$, consistent with the trivial inversion invariant $\nu_{\mathrm{inv}}=1$.

\section{Evolution of the energy gap in dimer phases}

\begin{figure}
\center
\includegraphics[width=0.5\textwidth]{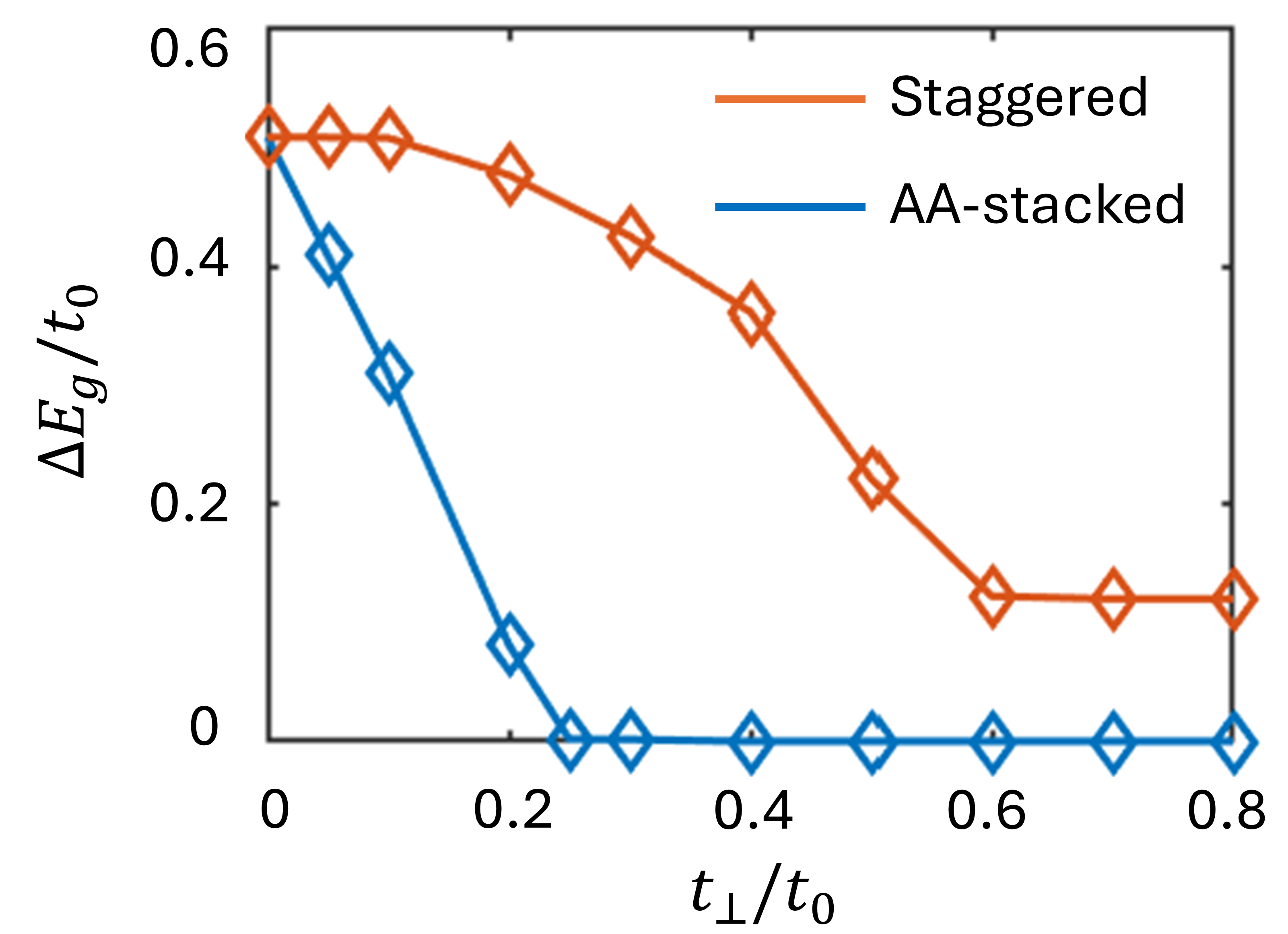}
\caption{Evolution of the energy gap as a function of interchain tunneling $t_{\perp}/t_0$ for the staggered- and AA-stacked dimer phases at $\alpha/t_0 = 0.8$.}
\label{supp_fig:1}
\end{figure}

Fig.~\ref{supp_fig:1} shows the dependence of the energy gap on the interchain tunneling $t_{\perp}/t_0$ for the staggered- and AA-stacked dimer phases at $\alpha/t_0 = 0.8$. In both cases the gap decreases as $t_{\perp}$ increases. The AA-stacked phase exhibits a rapid decline, with the gap becoming negligible around $t_{\perp}/t_0 \approx 0.25$. In contrast, the gap decreases slowly in the staggered-dimer phase, and beyond $t_{\perp}/t_0 \approx 0.6$ it approaches a nearly constant value.

\section{Evolution of the Energy with Interchain Displacement $\Delta x$}

\begin{figure}
\center
\includegraphics[width=1\textwidth]{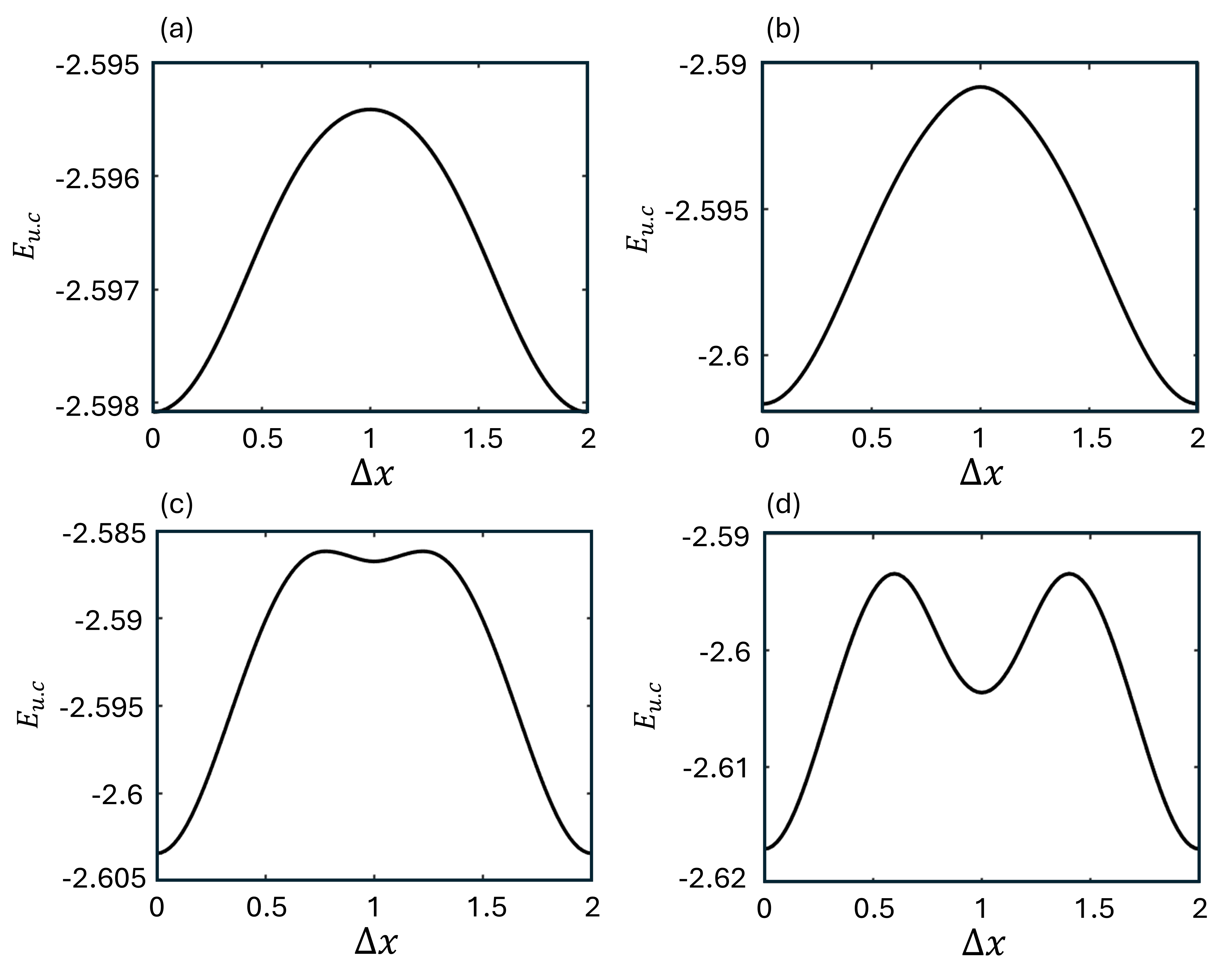}
\caption{Energy per unit cell as a function of interchain displacement $\Delta x$, for interchain hopping strengths (a) $t_\perp/t_0 = 0.1$, (b) $t_\perp/t_0 = 0.2$, (c) $t_\perp/t_0 = 0.3$, and (d) $t_\perp/t_0 = 0.4$. The displacements $\Delta x = 0, 2$ correspond to the staggered-dimer configuration, and $\Delta x = 1$ corresponds to the AA-stacked dimer phase.}
\label{supp_fig:2}
\end{figure}

Here we examine the dependence of the total energy per unit cell on the interchain displacement $\Delta x$ for an exponentially decaying interchain tunneling of the form
$t_\perp(\Delta x)=t_\perp\exp\!\left[-\gamma\left(\sqrt{\Delta x^2+1}-1\right)\right]$,
with $\gamma=5$.
Figure~\ref{supp_fig:2} shows the resulting energy profiles for several values of the interchain hopping strength $t_{\perp}/t_0$. For weak interchain tunneling, $t_{\perp}/t_0=0.1$ and $0.2$, the staggered-dimer configurations at $\Delta x=0$ and $2$ are the lowest-energy states. Starting from $\Delta x=0$, the energy increases monotonically as $\Delta x$ approaches $\Delta x=1$, corresponding to AA stacking, where it reaches a maximum; on increasing $\Delta x$ further, the energy decreases and reaches a minimum at $\Delta x=2$, the neighboring staggered-dimer configuration [Fig.~\ref{supp_fig:2}(a,b)].

For strong interchain tunneling, $t_{\perp}/t_0=0.3$ and $0.4$, the staggered-dimer configuration remains energetically favorable, but the energy profile develops additional features. Starting from $\Delta x=0$, the energy first increases and reaches a maximum before the AA-stacking configuration, then decreases toward $\Delta x=1$, followed by a corresponding increase on the opposite side of the AA-stacked configuration, and finally decreases toward the neighboring staggered-dimer configuration at $\Delta x=2$ [Fig.~\ref{supp_fig:2}(c,d)]. Thus, increasing $t_{\perp}$ introduces additional local maxima around the AA-stacked configuration while preserving the overall symmetry of the energy profile about $\Delta x=1$.

\section{Spatial Evolution of In-Gap States within a Moir\'e Supercell}

\begin{figure}
\center
\includegraphics[width=0.8\textwidth]{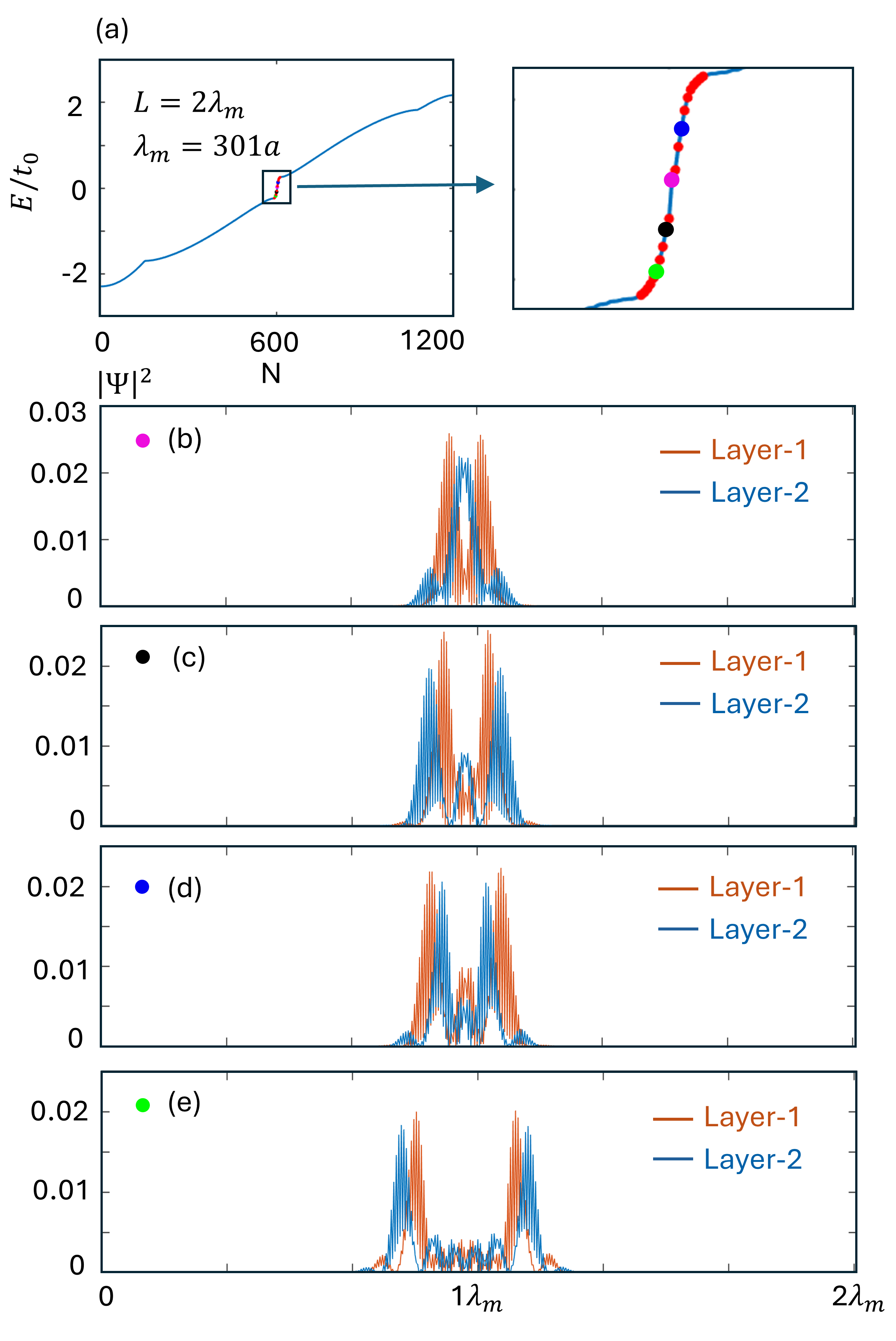}
\caption{Electronic band structure and real-space probability densities of the $L=2\lambda_m$ supermoir\'e phase. (a) Electronic band structure at $t_\perp/t_0 = 0.25$, extrapolated from a moir\'e wavelength $\lambda_m = 31a$ ($L=62a$) to $\lambda_m = 301a$ ($L=602a$). The spectrum is gapped, with colored markers indicating in-gap states localized between staggered-dimer regions. (b--e) Real-space probability density profiles of representative in-gap states, showing their localization near $x=\lambda_m$.}
\label{supp_fig:3}
\end{figure}

Here we examine the spatial evolution of the in-gap states in the extrapolated moir\'e supercell with $L=602a$ ($\lambda_m=301a$), obtained from the $L=62a$ ($\lambda_m=31a$) supercell at $t_{\perp}/t_0\approx0.25$. The corresponding band structure is shown in Fig.~\ref{supp_fig:3}(a), with representative spatial distributions of the in-gap states in Fig.~\ref{supp_fig:3}(b-e). The enlarged system yields a dense set of in-gap states within the bulk gap. States near the gap center are strongly localized around the AA-stacked region, where the local bilayer is nearly gapless at $t_{\perp}/t_0\approx0.25$; their wave functions have pronounced weight around the AA region and decay rapidly toward the neighboring stacking regions. As the energy moves away from the gap center, the states progressively extend into the surrounding regions with smaller local gaps. Near the bulk-gap boundaries, the states become more extended and acquire significant weight within the staggered-dimer regions, eventually merging with the delocalized bulk states. Thus, the localization length of the in-gap states varies continuously across the gap and is governed by the stacking-dependent local electronic structure of the supermoir\'e phase.

%\bibliography{supp_references.bib}

\providecommand{\latin}[1]{#1}
\makeatletter
\providecommand{\doi}
  {\begingroup\let\do\@makeother\dospecials
  \catcode`\{=1 \catcode`\}=2 \doi@aux}
\providecommand{\doi@aux}[1]{\endgroup\texttt{#1}}
\makeatother
\providecommand*\mcitethebibliography{\thebibliography}
\csname @ifundefined\endcsname{endmcitethebibliography}  {\let\endmcitethebibliography\endthebibliography}{}